\documentclass[10pt,onecolumn]{article}
\usepackage{amsmath,amssymb,amsfonts}
\usepackage{newtxtext}
\usepackage{newtxmath} 
\usepackage{framed,multirow}
\usepackage{comment}
\usepackage{soul}
\usepackage{booktabs}
\usepackage{algorithmic}
\usepackage[ruled,vlined]{algorithm2e}
\usepackage{graphicx}
\usepackage{textcomp}
\usepackage{placeins}
\usepackage{pifont} 
\usepackage{float}
\usepackage{array}
\usepackage{adjustbox}
\usepackage{bbm}
\usepackage{mathtools}
\usepackage{siunitx}
\usepackage{xcolor}
\usepackage{tikz}
\usetikzlibrary{matrix,calc}
\usepackage{latexsym}
\usepackage{gensymb}
\usepackage{url}
\usepackage{lipsum}
\usepackage{stfloats} 
\usepackage[utf8]{inputenc} 
\usepackage{longtable} 
\usepackage{nicefrac} 
\usepackage{microtype} 
\usepackage{nicematrix} 
\usepackage{makecell}

\usepackage[
    backend=biber,
    style=ieee,
    maxnames=6,
    minnames=1,
    url=false,      
    doi=false,      
    eprint= false,
]{biblatex}
\AtEveryBibitem{%
  \clearfield{issn}%
  \clearfield{isbn}%
}
\usepackage[margin=1in]{geometry} 
\usepackage{authblk} 

\usepackage{setspace}
\usepackage{titlesec}
\titleformat{\section}{\large\bfseries}{\thesection.}{0.5em}{}
\titleformat{\subsection}{\normalsize\bfseries}{\thesubsection.}{0.5em}{}

\usepackage[justification=justified, font=small,labelfont=bf]{caption}
\usepackage{subfig}
\usepackage{fix-cm}  
\SetKwInput{KwInput}{Input}                
\SetKwInput{KwOutput}{Output}              
\floatname{algorithm}{Algorithm}

\usepackage{hyperref}
\hypersetup{colorlinks=true, 
linkcolor=black, 
citecolor=blue,
filecolor=magenta, 
urlcolor=cyan, 
pdftitle={NeuroSSM}
}

\newcommand{\cmark}{\ding{51}}%
\newcommand{\xmark}{\ding{55}}%
\definecolor{bondiblue}{rgb}{0.0, 0.58, 0.71}
\definecolor{brightcerulean}{rgb}{0.11, 0.62, 0.74}
\newcommand*{\revhl}{\textcolor{black}}

\def\BibTeX{{\rm B\kern-.05em{\sc i\kern-.025em b}\kern-.08em
    T\kern-.1667em\lower.7ex\hbox{E}\kern-.125emX}}

\title{NeuroSSM: Multiscale Differential State-Space Modeling for Context-Aware fMRI Analysis}

\author[1,2]{Furkan Gen\c{c}}
\author[1,2]{Boran \.{I}smet Macun}
\author[1,2]{Sait Sarper \"{O}zaslan}
\author[1,2,3]{Emine U. Saritas}
\author[1,2,3,*]{\\Tolga \c{C}ukur}

\affil[1]{Department of Electrical and Electronics Engineering, Bilkent University, Ankara 06800, Turkey}
\affil[2]{National Magnetic Resonance Research Center (UMRAM), Bilkent University, Ankara 06800, Turkey}
\affil[3]{Department of Neuroscience, Bilkent University, Ankara 06800, Turkey}
\affil[*]{Corresponding author: cukur@ee.bilkent.edu.tr}

\date{} 

\begin{document}

  \begin{@twocolumnfalse}
    \maketitle
    
\begin{abstract}
\noindent Accurate fMRI analysis requires sensitivity to temporal structure across multiple scales, as BOLD signals encode cognitive processes that emerge from fast transient dynamics to slower, large-scale fluctuations. Existing deep learning (DL) approaches to temporal modeling face challenges in jointly capturing these dynamics over long fMRI time series. Among current DL models, transformers address long-range dependencies by explicitly modeling pairwise interactions through attention, but the associated quadratic computational cost limits effective integration of temporal dependencies across long fMRI sequences. Selective state-space models (SSMs) instead model long-range temporal dependencies implicitly through latent state evolution in a dynamical system, enabling efficient propagation of dependencies over time. However, recent SSM-based approaches for fMRI commonly operate on derived functional connectivity representations and employ single-scale temporal processing. These design choices constrain the ability to jointly represent fast transient dynamics and slower global trends within a single model. We propose NeuroSSM, a selective state-space architecture designed for end-to-end analysis of raw BOLD signals in fMRI time series. NeuroSSM addresses the above limitations through two complementary design components: a multiscale state-space backbone that captures fast and slow dynamics concurrently, and a parallel differencing branch that increases sensitivity to transient state changes. Experiments on clinical and non-clinical datasets demonstrate that NeuroSSM achieves competitive performance and efficiency against state-of-the-art fMRI analysis methods.

\vspace{1em}
\end{abstract}

    \vspace{1em} 
    
  \end{@twocolumnfalse}

\section{Introduction}
Functional magnetic resonance imaging (fMRI) non-invasively measures blood-oxygen-level-dependent (BOLD) signals that indirectly reflect neural activity distributed across the brain \autocite{hillman2014coupling}. Brain activity recorded during task-based \autocite{li2009review, venkataraman2009exploring, nishimoto2011} or resting-state \autocite{niu2021modeling, yeo2011organization} fMRI scans has therefore been widely analyzed to characterize cognitive processes associated with normal brain function and neurological disease \autocite{kong2019spatial, rajpoot2015functional}. Such cognitive processes involve neural activity across multiple temporal scales, ranging from fast transient dynamics to slower, globally coordinated trends \autocite{campbell2022monofractal}. However, fMRI does not observe these directly: neural activity is filtered through a slow and spatially heterogeneous hemodynamic response and sampled at relatively coarse temporal resolution \autocite{logothetis2001neurophysiological, handwerker2004variations}. As a result, observed BOLD signals exhibit long-range temporal dependencies and pronounced separation between fast and slow components \autocite{campbell2022monofractal}. Consequently, effective fMRI analysis requires models that can integrate information over long BOLD time series while remaining sensitive to brief state changes and robust under the limited sample sizes typical of neuroimaging studies.

Many existing deep learning (DL) approaches for fMRI analysis implicitly rely on modeling assumptions that conflict with these requirements. Convolutional, graph-based, and recurrent models have been commonly applied to fMRI \autocite{kawahara2017brainnetcnn, parisot2018disease, fan2020deep}, but they often operate on functional connectivity (FC) features derived from BOLD signals \autocite{mckeown1998independent, svensen2002ica, zhang2015resting, wang2019functional}. While FC representations reduce dimensionality and noise, they are typically computed using time-averaged correlation measures that assume stationarity and linear interactions, thereby collapsing temporal ordering \cite{lahaye2003functional}. This can obscure non-linear and time-varying dynamics present in the raw BOLD signals \autocite{hu2007nonlinear, ciuciu2012ScaleFreeAM}. As a result, such approaches trade temporal resolution for tractability, limiting their ability to capture temporal structure across extended time series.

Attention-based transformer architectures have recently been adopted to address long-range temporal dependencies in fMRI by explicitly modeling interactions across time \autocite{malkiel2021pre, peng2024gbt, kim2021learning, bedel2023bolt}. In these models, each time point is treated as a discrete token, and self-attention captures pairwise dependencies based on token similarity. However, this formulation introduces limitations that are particularly pronounced for fMRI analysis. The quadratic computational and memory costs of self-attention restrict effective integration across long BOLD sequences \autocite{vaswani2017attention}. More fundamentally, higher-order and multiscale temporal structure must be inferred indirectly from pairwise interactions rather than being explicitly represented, a limitation that persists even when attention mechanisms are made more computationally efficient through sparsification or approximation \autocite{zhang2022diffusion,bedel2023bolt}.

Within the current landscape of fMRI analysis models, jointly capturing latent temporal dynamics while maintaining stable and computationally practical processing over long time series remains challenging. State-space models (SSMs) provide a principled framework for modeling temporal structure by describing signal evolution through latent state dynamics \autocite{gu2022efficientlymodelinglongsequences}. Applying SSMs to fMRI, however, raises additional design questions regarding how temporal structure should be represented across multiple time scales. Recent SSM-based studies in fMRI have explored specific design choices \autocite{behrouz2024brainmamba, wei2025hierarchicalspatiotemporalstatespacemodeling, deng2025causalmamba}, often emphasizing connectivity-based representations and single-scale temporal modeling. While these approaches demonstrate the suitability of SSMs for fMRI analysis, they do not explicitly address how fast transient dynamics and slower global trends can be jointly represented within a single model.

Here we introduce NeuroSSM, a selective state-space model designed for high-fidelity fMRI analysis through two complementary design components. First, NeuroSSM employs a multiscale state-space backbone that processes BOLD signals at both full temporal resolution and coarser scales, enabling concurrent modeling of fast transient dynamics and slower global trends. Second, a parallel differencing branch incorporates first-order temporal derivatives of the BOLD signal into the state-space formulation, increasing sensitivity to transient state changes while preserving long-range temporal integration. Together, these components allow NeuroSSM to encode temporal structure across multiple scales while maintaining linear computational cost with respect to sequence length.

We evaluate NeuroSSM on two public datasets: the Human Connectome Project (HCP) \autocite{van2013wu}, including resting-state and task-based fMRI data, and the Parkinson’s Progression Markers Initiative (PPMI) \autocite{marek2018ppmi}, focused on distinguishing Parkinson’s disease patients from controls. Across both datasets, NeuroSSM demonstrates competitive predictive performance relative to classical DL, transformer, and SSM baselines. Ablation studies further confirm the importance of both the multiscale backbone and the differencing branch to overall model performance.

\section*{Contributions}
\begin{itemize}
\item We propose NeuroSSM, a selective state-space architecture for end-to-end analysis of raw fMRI BOLD time series, with computational cost that scales linearly with sequence length.
\item We introduce a multiscale state-space backbone that enables joint modeling of fast transient dynamics and slower global trends in BOLD signals within a single model.
\item We incorporate a parallel differencing branch into the state-space formulation to increase sensitivity to rapid changes in brain state.
\item We evaluate NeuroSSM on large-scale public fMRI datasets spanning clinical and non-clinical settings, demonstrating performance competitive with state-of-the-art fMRI analysis methods.
\end{itemize}

\section{Related Work}
\subsection{Earlier DL Architectures} 
In the early adoption of DL for fMRI, models typically addressed the high-dimensional and noisy nature of raw BOLD signals through dimensionality reduction. A common strategy has been to pre-process the BOLD time series into FC matrices, which capture pairwise linear relationships between brain regions \autocite{zhang2015resting, wang2019functional}. These matrices have been subsequently used as input to DL models, such as convolutional neural networks (CNNs) \autocite{kawahara2017brainnetcnn} or graph neural networks (GNNs) \autocite{parisot2018disease, kim2020understanding, kan2022fbnetgen}, to extract spatial information from the connectivity structure. While lowering computational load, FC features primarily encode static, first-order interactions, consequently overlooking the complex, non-linear dynamics inherent in BOLD signals \cite{lahaye2003functional, hu2007nonlinear}. As the field pivoted toward explicit temporal sequence modeling, recurrent neural networks (RNNs) were adopted as a natural fit \autocite{yang2018aRobustDNN, li2018lstm}. However, performance of RNNs is often hampered by the vanishing or exploding gradient problem in practice, hindering their ability to propagate information effectively across long ranges \autocite{kerg2020untangling}. Moreover, their sequential, non-parallelizable nature increases training latency for lengthy fMRI scans \autocite{pascanu2013difficulty}.

\subsection{Transformer Architectures} 
In recent years, self-attention has been presented as a powerful mechanism to capture contextual interactions across time \autocite{vaswani2017attention, nguyen2020attend, he2024spatiotemporal}. Transformer models treat the BOLD signal as a sequence of tokens, where each token represents the multivariate activity across all brain regions at a single time point (TR). However, vanilla self-attention extracts dependencies based on pair-wise similarity between tokens, limiting the capacity to integrate contextual interactions across extended sequences and capture higher-order interactions characteristic of neural activity \autocite{zhang2022diffusion}. Furthermore, vanilla self-attention introduces quadratic complexity with respect to sequence length, which presents a significant computational bottleneck for long fMRI scans \autocite{vaswani2017attention}.

Efficient transformer variants have later emerged, including dividing the full scan into overlapping windows to achieve localized attention \autocite{malkiel2021pre}; employing a parallel hybrid network to extract local and global representations with interactive fusion \autocite{zhao2022ift}; or using low-rank decompositions to approximate the full attention matrix \autocite{peng2024gbt}. While these methods achieve near-linear scaling in run time and memory, they introduce a context-efficiency trade-off: windowing restricts information propagation across partitions, down-sampling sacrifices local temporal sensitivity, and low-rank approximation risks underfitting by discarding subtle high-rank temporal patterns \autocite{bedel2023bolt, tian2025transformerforfMRI}. These approximations often undermine the original promise of global context integration, particularly missing very long-range interactions intrinsic to fMRI data \autocite{campbell2022monofractal}, and require careful re-tuning of customized approximation strategies for specific fMRI datasets.

\subsection{State-Space Models (SSMs)}
SSMs provide a promising solution to the sequence modeling dilemma by modeling sequences as discretized linear dynamical systems \autocite{gu2022efficientlymodelinglongsequences}. They are inherently well-suited for modeling very long temporal dependencies, leading several recent studies to adapt SSMs to fMRI analysis. However, the existing SSM architectures for fMRI possess common limitations that prevent them from fully exploiting the hierarchical nature of BOLD signals. One characteristic of existing methods is their reliance on derived FC features, which can circumvent direct temporal modeling of raw BOLD signals \autocite{behrouz2024brainmamba, wei2025hierarchicalspatiotemporalstatespacemodeling}. On the other hand, methods that operate on BOLD signals commonly perform processing at a single temporal scale \autocite{deng2025causalmamba}. Note that the efficiency of SSMs relies on compressing contextual representations during sequence aggregation, but such single-scale compression forces a suboptimal compromise that cannot simultaneously focus on distinct temporal frequencies. In turn, single-scale SSMs can struggle to simultaneously resolve rapid fluctuations and global trends.  

NeuroSSM distinguishes itself from existing SSMs by discarding both FC features and single-scale processing. It introduces a multiscale backbone coupled with a temporal differencing pathway, jointly reshaping the input sequence across temporal resolutions to capture both rapid transients and long-range trends. This unified architecture enables end-to-end modeling of hierarchical brain dynamics while maintaining strict linear complexity.

\section{Theory}
\label{sec:Theory}
For multivariate analysis of 4D fMRI data, NeuroSSM first parcellates the brain into $N$ regions of interest (ROIs) using an anatomical atlas. Within each ROI, the voxel-wise BOLD signals are averaged to produce a single time series, which is then z-scored to zero mean and unit variance. We use $K$ temporal scales with step sizes $\boldsymbol{\tau}=(\tau_1,\dots,\tau_K)$. For each scale $k$, temporal reshaping yields a sequence length $T_k=T/\tau_k$ and a feature dimension $d_k=\tau_k N$. Our model learns a mapping from these regional BOLD responses to target labels $y$ (e.g. subject diagnosis or task conditions), treating each time-indexed vector as a token in a temporal sequence.

\begin{figure*}[t]
  \centering
\includegraphics[width=0.99\linewidth]{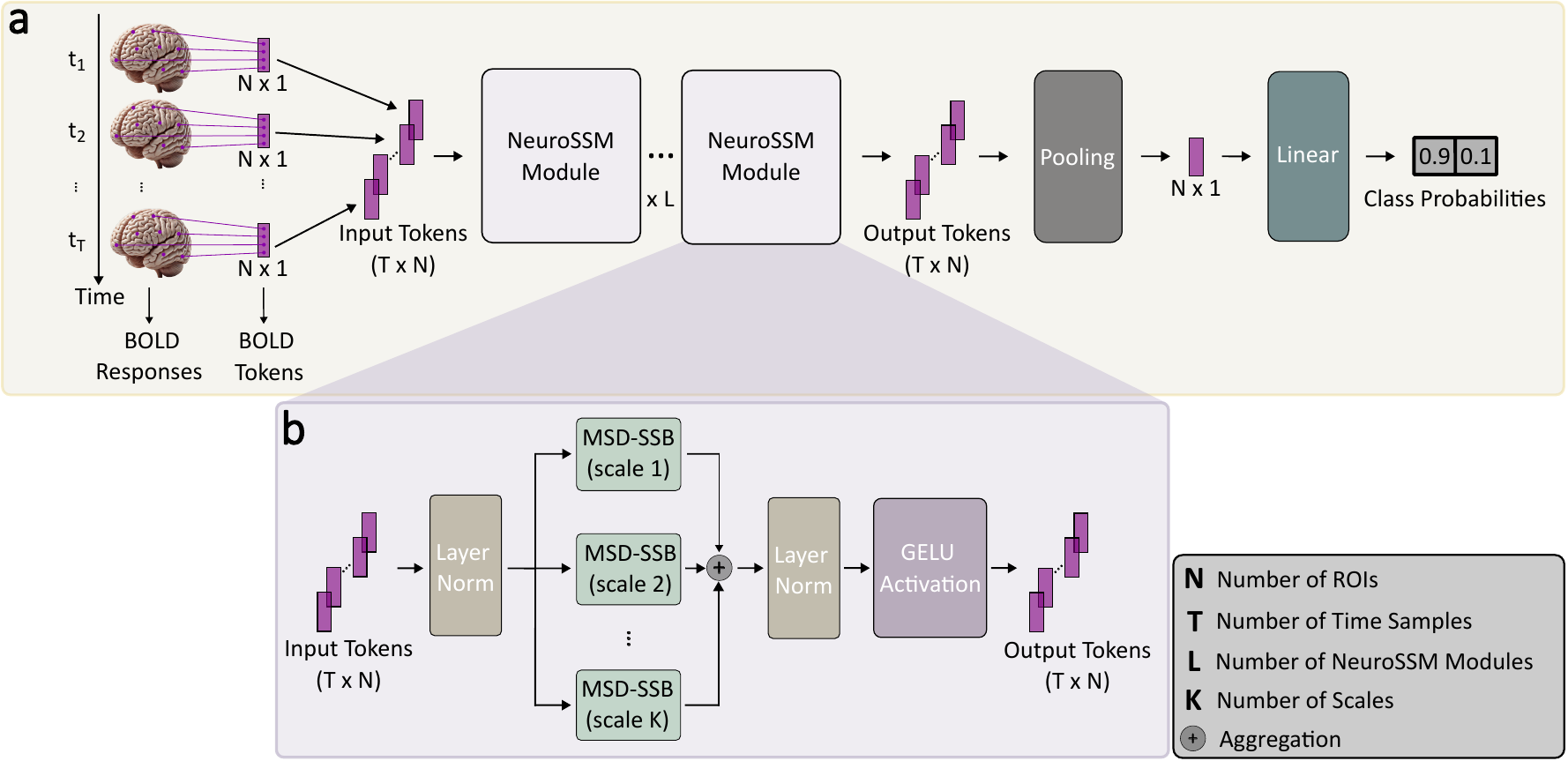}  
  \caption{\textbf{Overview of the NeuroSSM framework.} \textbf{(a)} The high-level processing pipeline. Region-of-interest (ROI) BOLD signals are extracted from 4D fMRI scans to form a multivariate time series of dimension $T \times N$, where $T$ is the sequence length and $N$ is the number of ROIs. This sequence is processed by a stack of $L$ NeuroSSM modules, which extract hierarchical spatiotemporal features while preserving the sequence length. The final output sequence is aggregated via temporal pooling into a single $N$-dimensional representation, which is then mapped by a linear classifier to task-specific class probabilities (e.g., probability of disease presence or absence). \textbf{(b)} Internal structure of a NeuroSSM block. The input sequence is normalized and distributed to a bank of Multiscale Differential State-Space Blocks (MSD-SSB) operating in parallel across $K$ temporal scales. Each scale $k$ processes the sequence with a distinct step size $\tau_k$, enabling the simultaneous capture of fine-grained transients and slow-varying global trends. The multi-scale outputs are summed, normalized, and activated via GeLU before propagating to subsequent layers.}
\label{fig:multiscale}
\end{figure*}

Let $\mathbf{x}=\bigl(x^{(0)},\dots,x^{(T-1)}\bigr)\in\mathbb{R}^{T\times N}$ denote the $N$-variate BOLD time-series extracted from an fMRI scan of duration $T$. We treat each vector $x^{(t)}$ as a length-$N$ \emph{token} and stacked them into the sequence
\[
\mathbf{B}_{0}=\bigl(x^{(0)},x^{(1)},\dots,x^{(T-1)}\bigr)\in\mathbb{R}^{T\times N}.
\]
This raw sequence is processed by a cascade of $L$ \emph{NeuroSSM} modules followed by a task-specific head. Each \emph{NeuroSSM} module first normalizes its input, then applies a bank of \emph{multiscale differential state space blocks} (MSD-SSB). A schematic overview of this pipeline is provided in Figure~\ref{fig:multiscale}. 

\begin{figure*}[t]
  \centering
\includegraphics[width=0.99\linewidth]{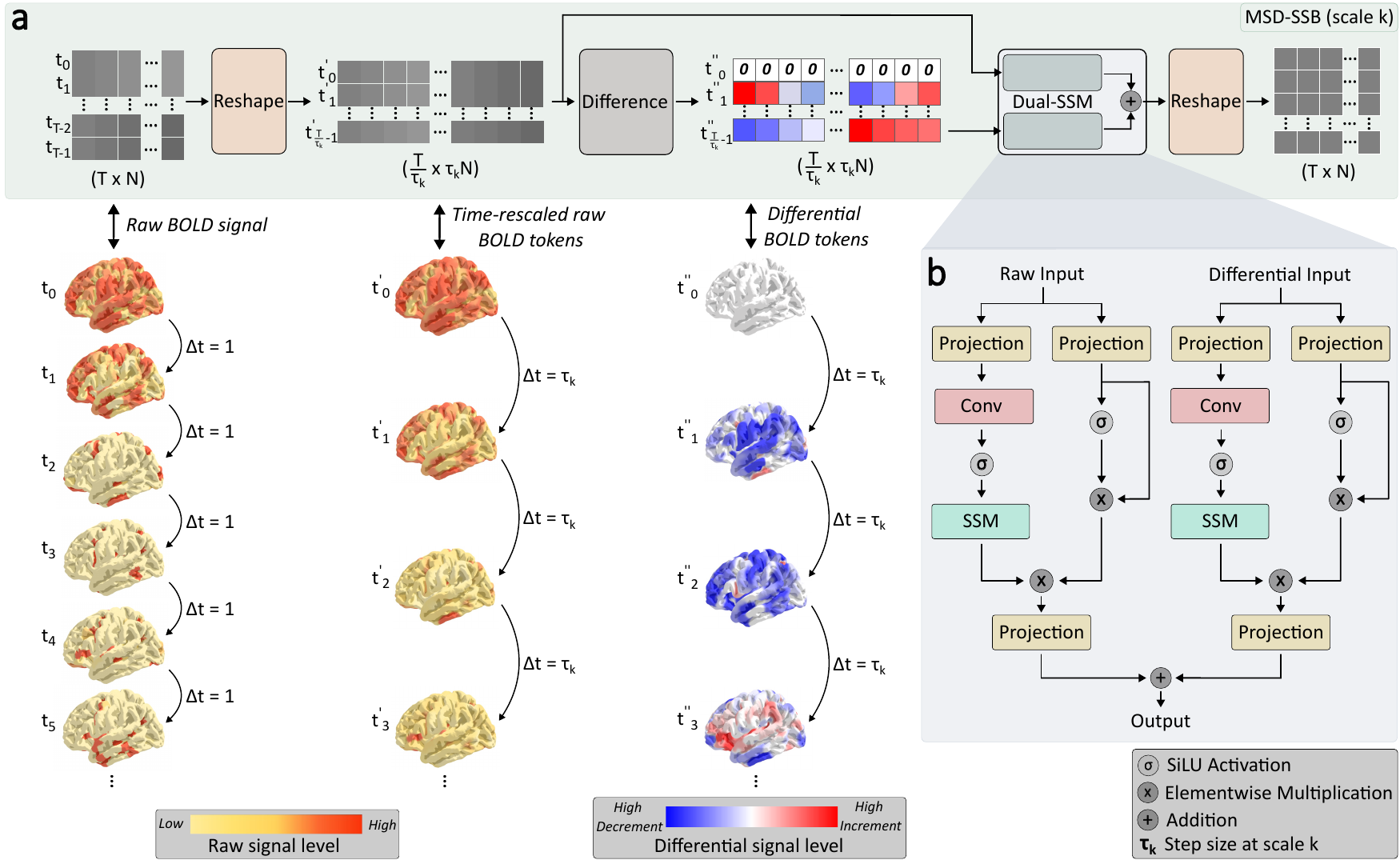}
\caption{\textbf{Detailed structure of MSD-SSB (Multiscale Differential State-Space Block).} \textbf{(a)} The $T \times N$ input sequence is temporally reshaped at scale $k$ with a step size $\tau_k$ into a representation of size $\lfloor T/\tau_k \rfloor \times (\tau_k \cdot N)$. Two parallel input streams are then generated: (i) the $\tau_k$-rescaled BOLD sequence and (ii) its first-order temporal differences, emphasizing rapid state transitions. These two sequences are fed into a Dual-SSM block. The outputs are summed via residual connection and reshaped back to the original $T \times N$ dimensions for layer propagation. \textbf{(b)} The Dual-SSM block processes the raw and differential streams concurrently using parameter-shared selective state-space kernels. Each stream input undergoes a main path involving a depth-wise convolutional layer, SiLU activation, and the SSM to capture long-range dependencies, and a second path involving a nonlinear projection and SiLU activation, functioning as a multiplicative gating mechanism. The outputs of raw and differential streams are additively fused.}
    \label{fig:difference}
\end{figure*}

\subsection{NeuroSSM Module}
\label{subsec:NeuroSSM}
Each \emph{NeuroSSM} module receives a sequence $\mathbf{B}_{\ell-1}\in\mathbb{R}^{T_{\ell-1}\times N}$ and outputs $\mathbf{B}_{\ell}\in\mathbb{R}^{T_{\ell}\times N}$ ($\ell=1\dots L$). For simplicity, we drop the layer index and analyze a single block here. The module first balances channel statistics with an \emph{input layer normalization}. For a sequence tensor $\mathbf{B}\in\mathbb{R}^{T\times d}$ (length $T$, feature dimension $d$), this is achieved via:
\begin{equation}
\widetilde{\mathbf{B}}
=\mathrm{LN}_{\mathrm{in}}(\mathbf{B})
=\gamma_{\mathrm{in}}\odot\frac{\mathbf{B}-\mu(\mathbf{B})}{\sqrt{\sigma^{2}(\mathbf{B})+\varepsilon}}
+\beta_{\mathrm{in}},
\label{eq:layernorm}
\end{equation}
where
\[
\mu(\mathbf{B})=\frac{1}{d}\sum_{c=1}^{d}\mathbf{B}_{c},\quad
\sigma^{2}(\mathbf{B})=\frac{1}{d}\sum_{c=1}^{d}\bigl(\mathbf{B}_{c}-\mu(\mathbf{B})\bigr)^{2},
\]
and $\gamma_{\mathrm{in}}, \beta_{\mathrm{in}}\in\mathbb{R}^{d}$ are learnable gain and bias vectors, with $\varepsilon>0$ for numerical stability. The resulting $\widetilde{\mathbf{B}}$ is then forwarded to the multiscale processing stages.

\subsubsection{Multiscale Differential State Space Block (MSD-SSB)}
\label{subsubsection:MSD-SSB}
MSD-SSB is designed to capture dynamics across multiple temporal hierarchies. For each scale $k=1,\dots,K$, we first time-rescale the input $\widetilde{\mathbf{B}}\in\mathbb{R}^{T\times N}$ by a step size $\tau_k$, grouping every $\tau_k$ consecutive tokens into one rescaled token:
\begin{equation}
\widetilde{\mathbf{B}}^{\langle k\rangle}_{t}
\;=\;\operatorname{vec}\bigl(\widetilde{\mathbf{B}}_{\,t\,\tau_k : t\,\tau_k + \tau_k - 1}\bigr),
\quad
t=0,1,\dots,\frac{T}{\tau_k}-1,
\end{equation}
so that the rescaled sequence tensor is defined as
\[
\widetilde{\mathbf{B}}^{\langle k\rangle}
\;\in\;\mathbb{R}^{\frac{T}{\tau_k}\times (\tau_k N)}.
\]
Here $\tau_k$ represents the time-rescaling step at scale $k$; for example, choosing $\tau_k = k$ yields the set of steps $\boldsymbol{\tau}=(1,2,3)$ when $K=3$.

Concurrently, at each rescaled level, we compute a first-order temporal difference to heighten sensitivity to transient activity by explicitly injecting rate of signal change:
\begin{equation}
\Delta\widetilde{\mathbf{B}}^{\langle k\rangle}_{0} = \mathbf{0},
\quad
\Delta\widetilde{\mathbf{B}}^{\langle k\rangle}_{t}
= \widetilde{\mathbf{B}}^{\langle k\rangle}_{t}
- \widetilde{\mathbf{B}}^{\langle k\rangle}_{\,t-1},
\quad
t = 1,\dots,\frac{T}{\tau_k}-1.
\label{eq:diffStream}
\end{equation}

The zero vector initialized at $t=0$ prevents spurious start-up artifacts, and the backward difference emphasizes frame-to-frame changes without altering the sequence length or dimensionality. Both the time-rescaled signal $\widetilde{\mathbf{B}}^{\langle k\rangle}$ and its temporal differential $\Delta\widetilde{\mathbf{B}}^{\langle k\rangle}$ are then processed in parallel by our dual-stream state-space blocks (Dual-SSM), producing $\mathbf{U}^{\langle k\rangle}$ and $\mathbf{V}^{\langle k\rangle}$, whose outputs are fused and reshaped back to the original temporal resolution in the subsequent fusion stage. Figure \ref{fig:difference} depicts the structure of the MSD-SSB equipped with Dual-SSM. 

\subsubsection{Dual-Stream State-Space Module (Dual-SSM)}
The Dual-SSM processes both the \emph{time-rescaled stream} and its \emph{temporal difference} in parallel using two specialized state-space kernels, $\mathrm{SSM}_k^{(\mathrm{rescaled})}$ and $\mathrm{SSM}_k^{(\mathrm{diff})}$.
For each scale $k$, we first linearly project the rescaled and differential inputs using learnable mappings $s_{\mathrm{rescaled}}^{(k)}$ and $s_{\mathrm{diff}}^{(k)}$, respectively:
\begin{equation}
X^{\langle k\rangle}
= s_{\mathrm{rescaled}}^{(k)}\bigl(\widetilde{\mathbf{B}}^{\langle k\rangle}\bigr),
\quad
D^{\langle k\rangle}
= s_{\mathrm{diff}}^{(k)}\bigl(\Delta\widetilde{\mathbf{B}}^{\langle k\rangle}\bigr),
\end{equation}

Here $s_{\mathrm{rescaled}}^{(k)}$ and $s_{\mathrm{diff}}^{(k)}$ map from $d_k=\tau_k N$ channels to an internal width $d_{\mathrm{inner},k}=\mathrm{expand}\cdot d_k$ (implemented via a gated linear projection). 
We then apply depth-wise 1D convolution (denoted by $*$) with kernel width $d_{\mathrm{conv}}$ and per-scale kernels $C_{\mathrm{rescaled}}^{(k)}$ and $C_{\mathrm{diff}}^{(k)}$,
\begin{equation}
\overline{X}^{\langle k\rangle}
= X^{\langle k\rangle} * C_{\mathrm{rescaled}}^{(k)},
\quad
\overline{D}^{\langle k\rangle}
= D^{\langle k\rangle} * C_{\mathrm{diff}}^{(k)}.
\end{equation}

These convolved outputs are then fed into the two specialized selective state-space kernels: $\mathrm{SSM}_k^{(\mathrm{rescaled})}$ for the rescaled stream and $\mathrm{SSM}_k^{(\mathrm{diff})}$ for the differential stream:
\begin{equation}
\mathbf{U}^{\langle k\rangle}
= \mathrm{SSM}_k^{(\mathrm{rescaled})}\bigl(\overline{X}^{\langle k\rangle}\bigr),
\quad
\mathbf{V}^{\langle k\rangle}
= \mathrm{SSM}_k^{(\mathrm{diff})}\bigl(\overline{D}^{\langle k\rangle}\bigr).
\end{equation}

\noindent
Each kernel $\mathrm{SSM}_k$ maintains learned diagonal state coefficients
$\Lambda\in\mathbb{R}^{d_{\mathrm{inner},k}\times d_{\mathrm{state}}}$,
where $d_{\mathrm{state}}$ is the per-channel state sizes, and three compact MLPs
$s_{\delta},s_{\beta},s_{\gamma}$ that map the convolved token at time $t$,
$\overline{r}^{\langle k\rangle}_{t}$ (i.e., $\overline{X}^{\langle k\rangle}_{t}$ for
$\mathrm{SSM}_k^{(\mathrm{rescaled})}$ or $\overline{D}^{\langle k\rangle}_{t}$ for
$\mathrm{SSM}_k^{(\mathrm{diff})}$), to content-aware selection parameters:
\begin{equation}
\delta^{(t)} = s_{\delta}(\overline{r}^{\langle k\rangle}_{t}),\quad
\beta^{(t)} = s_{\beta}(\overline{r}^{\langle k\rangle}_{t}),\quad
\gamma^{(t)} = s_{\gamma}(\overline{r}^{\langle k\rangle}_{t}),
\end{equation}
with $\delta^{(t)}>0$. These determine the discrete update coefficients for the linear recurrence:
\begin{equation}
A^{(t)} = \exp\bigl(\delta^{(t)}\Lambda\bigr),
\quad
P^{(t)} = \bigl(\delta^{(t)}\Lambda\bigr)^{-1}
 \bigl(e^{\delta^{(t)}\Lambda}-I\bigr)\,\delta^{(t)} \beta^{(t)},
\label{eq:ssm-discrete-fixed}
\end{equation}
which drives the linear recurrence relation:
\begin{equation}
h^{(t)} = A^{(t)}\,h^{(t-1)} + P^{(t)}\,\overline{r}^{\langle k\rangle}_{t},
\quad
u^{(t)} = \gamma^{(t)} \odot h^{(t)},
\label{eq:ssm-recur-fixed}
\end{equation}
where $\odot$ denotes element-wise multiplication. In parallel, another lightweight MLP $s_z$ produces a gate vector:
\begin{equation}
z^{(t)} = s_{z}(\overline{r}^{\langle k\rangle}_{t}).
\end{equation}
and we apply the tri-gate nonlinearity with a final linear projection $s_{\mathrm{out}}$:
\begin{equation}
y^{(t)} = u^{(t)} \odot \bigl(z^{(t)} \odot \sigma(z^{(t)})\bigr),
\quad
o^{(t)} = s_{\mathrm{out}} \bigl(y^{(t)}\bigr).
\label{eq:ssm-output-fixed}
\end{equation}

Because $\Lambda$ is diagonal (stored as per-channel state coefficients), the recurrence in
\eqref{eq:ssm-discrete-fixed} and \eqref{eq:ssm-recur-fixed} is element-wise across the
$d_{\mathrm{inner},k}$ channels and $d_{\mathrm{state}}$ state dimensions, enabling a fused
\emph{selective-scan} implementation with $\mathcal{O}\!\left(d_{\mathrm{inner},k}\,d_{\mathrm{state}}\right)$
cost per time step (i.e., linear in $d_{\mathrm{inner},k}$ for fixed $d_{\mathrm{state}}$).
The final projection in \eqref{eq:ssm-output-fixed} are applied after the scan. Finally, the two streams of the Dual-SSM are summed and reshaped back to the original sequence length:
\begin{equation}
\mathbf{Z}^{\langle k\rangle}
= \bigl(\mathbf{U}^{\langle k\rangle} + \mathbf{V}^{\langle k\rangle}\bigr)
 \;\overset{\mathrm{reshape}}{\longrightarrow}\;\mathbb{R}^{T\times N}.
\end{equation}

\subsubsection{Multiscale Fusion and Prediction}
After Dual-SSM processing, the outputs from each scale are aligned to the common $T\times N$ shape and are fused across scales via a parameter-free residual sum:
\begin{equation}
\mathbf{Z}
 =\sum_{k=1}^{K}\mathbf{Z}^{\langle k\rangle}.
\label{eq:fusion}
\end{equation}
This simple aggregation preserves information from every scale while avoiding the extra parameters and computational overhead of learned (e.g., attention-based) fusion schemes.

The fused sequence still carries scale-specific statistics; therefore, we apply a second layer normalization, $\mathrm{LN}_{\text{out}}(\cdot)$, using the same functional form as in ~\eqref{eq:layernorm} but with its own learnable parameters. A Gaussian-error linear unit (GELU) is then applied element-wise,
\begin{equation}
\mathbf{B}_{\mathrm{out}}
 =\phi\bigl(\mathrm{LN}_{\text{out}}(\mathbf{Z})\bigr),
\quad
\phi(z)=\tfrac{1}{2}z\bigl[1+\operatorname{erf}(z/\sqrt{2})\bigr],
\label{eq:postAct}
\end{equation}
providing smooth activation and improved gradient flow. Stacking $L$ of these \emph{NeuroSSM} modules yields the deep representation $\mathbf{B}_{L}$, which retains both local dynamics and long-range dependencies. To obtain a fixed-size representation, we perform global mean pooling over time,
\begin{equation}
\mathbf{g}= \frac{1}{T}\sum_{t=0}^{T-1}\mathbf{B}_{L,t},
\end{equation}
which captures the overall temporal context without adding extra parameters. The pooled representation $\mathbf{g}$ is then passed through a linear classifier for final prediction.

\subsection{Model complexity}
\label{complexity}
The computational cost of a sequence model fundamentally depends on how often an $N$-dimensional token interacts with the remaining $T-1$ tokens in the sequence. Among classical DL architectures, \emph{RNNs} update an $N \times N$ transition matrix at every time-step, which yields $O(N^{2}T)$ time and $O(N^{2})$ memory complexity \autocite{dvornek2017identifying,zhao20203d,xing2019dynamic}. A 1D \emph{CNN} with kernel width $k_{\mathrm{cnn}}$ performs $k_{\mathrm{cnn}}$ such matrix-vector products, resulting in $O(k_{\mathrm{cnn}}N^{2}T)$ time and $O(k_{\mathrm{cnn}}N^{2})$ memory complexity. \emph{Vanilla transformers} compute dot-products between every pair of the $T$ queries and keys; thus, multi-head self-attention scales as $O(NT^{2})$ in both time and memory \autocite{vaswani2017attention,nguyen2020attend,zhao2022ift}. Meanwhile, \emph{efficient transformers} restrict attentional interactions to local windows or sparse patterns. For example, methods utilizing temporal windowing or local attention segment the sequence into overlapping windows of length $W$ and restrict self-attention interactions to tokens within that window \autocite{bedel2023bolt}. The total computational cost across the sequence is thereby reduced to $O(NTW)$, achieving linear scaling with respect to the sequence length $T$ while introducing $W$ as a new, performance-critical hyperparameter that dictates the maximum context length \autocite{bedel2023bolt}. Hence as $W$ is grown to capture long-range context more effectively, the efficiency benefits are compromised to go back to quadratic complexity. 

In comparison, \emph{NeuroSSM} eliminates attention entirely by using $K$ parallel selective SSM streams.
At scale $k$, temporal reshaping produces a sequence of length $T_k=T/\tau_k$ with token width $d_k=\tau_k N$. Dual-SSM expands this width to $d_{\mathrm{inner},k}=\mathrm{expand}\cdot d_k$ and maintains a per-channel latent state of size $d_{\mathrm{state}}$. Since the state-transition is diagonal, the selective-scan update is computed element-wise across channels, costing $\mathcal{O}(d_{\mathrm{inner},k} d_{\mathrm{state}})$ per time step and $\mathcal{O}(T_k d_{\mathrm{inner},k} d_{\mathrm{state}})$ per scale. Substituting $T_k=T/\tau_k$ and $d_{\mathrm{inner},k}=\mathrm{expand}\cdot\tau_k N$ yields $\mathcal{O}(\mathrm{expand}\cdot N\cdot d_{\mathrm{state}}\cdot T)$ time per scale, and thus $\mathcal{O}(K \cdot \mathrm{expand}\cdot N\cdot d_{\mathrm{state}}\cdot T)$ time overall. This corresponds to $\mathcal{O}(d_{\mathrm{inner},k}d_{\mathrm{state}})$ memory per scale (linear in $N$ for fixed $d_{\mathrm{state}}$).
The gating and final projection are applied after the scan and do not change the linear scaling in $T$. In our implementation, $K=3$, $\mathrm{expand}=3$, and $d_{\mathrm{state}}=2$ are small constants; thus the dominant time cost scales as $\mathcal{O}(N T)$ and the memory scales as $\mathcal{O}(N)$, up to the fixed factor $K\cdot \mathrm{expand}\cdot d_{\mathrm{state}}$.

\section{Methods}

\subsection{Datasets} \label{Dataset}
Demonstrations were performed on the HCP S1200 \footnote{\url{https://db.humanconnectome.org}} \autocite{van2013wu} and PPMI releases \autocite{marek2018ppmi}. Within HCP S1200, resting-state fMRI data (HCP-Rest) were analyzed for gender detection, whereas task-based fMRI data (HCP-Task) were analyzed for cognitive task detection. In PPMI, resting-state fMRI data were analyzed to detect Parkinson’s disease (PD). All data were preprocessed via rigid alignment to structural scans, normalization to MNI space, ROI parcellation, and z-scoring of the time course in each ROI prior to modeling.

\textbf{HCP-Rest:} The first session of preprocessed resting-state fMRI data were used \autocite{glasser2013minimal}. Scans shorter than 1200 time samples were discarded. The analyzed dataset consisted of 871 scans, with 484 female and 387 male subjects. Parcellations were performed via the Schaefer atlas, yielding 400 ROIs \autocite{schaefer2017schaefer}.

\textbf{HCP-Task:} The first session of preprocessed task-based fMRI data were used \autocite{glasser2013minimal}. Scans shorter than the canonical lengths for each of the seven cognitive tasks (emotion, relational, gambling, language, social, motor, and working memory) were discarded. The analyzed dataset consisted of 427 scans from 61 subjects. Parcellations were performed via the Schaefer atlas, yielding 400 ROIs \autocite{schaefer2017schaefer}.

\textbf{PPMI:} Preprocessed resting-state fMRI data were used \autocite{marek2018ppmi}. A binary disease detection task was used where the positive class comprised subjects with 
Parkinson's disease, and the negative class comprised neurologically healthy controls. Prodromal, SWEDD, and all genetic cohort arms were excluded. For subjects with multiple resting-state runs, only the first available scan was retained. The analyzed dataset contained 114 subjects, with 99 PD and 15 control cases. Parcellations were performed via the AAL3v2 atlas, yielding 166 ROIs \autocite{rolls2020automated}.

\subsection{Experimental Setup} \label{ExperimentSettings}
All models were implemented using the PyTorch framework on a single Nvidia RTX 5090 GPU. Models were trained to minimize the categorical cross-entropy loss via Adam optimizer, using 20 epochs, a mini-batch size of 32, and a weight decay of $4 \times 10^{-5}$, unless otherwise noted.

For all datasets, we adopted a stratified cross-validation procedure for assessing model performance. Data were first partitioned into a development set (80\%) and a held-out test set (20\%), which maintained balance between different classes. Within the development set, hyperparameters were tuned using 5-fold group cross-validation. In each inner fold, the model was trained on approximately 64\% of the samples and validated on the remaining 16\%. After selecting hyperparameters based on mean validation performance, a final model was trained on the full development set and evaluated once on the held-out test set. Subject identities were strictly kept disjoint across training, validation, and test sets throughout all experiments.

To analyze learning efficiency, we constructed learning curves on each dataset. For a given dataset, we subsampled subjects from the development set to use only a fraction $S \in \{5, 10, 20, 50, 100\}\%$ of the available training subjects, while keeping the held-out test set fixed. For each $S$, models were retrained from scratch on the corresponding subset of the development set and evaluated on the test set. 

To enhance training efficiency and introduce stochasticity, random temporal cropping was applied only during training: 600 time points for HCP-Rest, 150 for HCP-Task, and 100 for PPMI. Test sequences were preserved at their full original length. For fair comparison, all models shared identical subject-wise splits and training samples. Experiments were conducted with five random seeds, and results were averaged to improve reliability, with seeds held constant across models. Model performance was quantified via accuracy, F1 score, and AUC. Statistical significance of performance differences between NeuroSSM and baseline methods was assessed via non-parametric Wilcoxon signed-rank tests.

\subsection{Implementation Details}\label{configuration}
The core NeuroSSM model employed a single block layer ($L=1$) across all experiments. The input at each time step was the parcellated ROI vector of dimension $N$ (e.g., $N=400$ for HCP and $N=166$ for PPMI). This input vector was passed through the NeuroSSM block, which consisted of: an initial Layer Normalization, a bank of Multiscale Differential State-Space Blocks (MSD-SSB) operating at $K=3$ temporal scales with step sizes $\boldsymbol{\tau}=(1,2,3)$, a second Layer Normalization, and a GeLU activation.

For each scale $\tau_k$, the MSD-SSB first performed $\tau_k$-fold temporal rescaling and computed a first-order temporal difference stream. Both streams were then processed in parallel by the Dual-Stream State-Space Module (Dual-SSM). The Dual-SSM contained two identical single-input/single-output state-space kernels (one for the rescaled input and one for the differential input), which shared parameterization to maintain efficiency. These kernels were configured with a latent state size $d_{\mathrm{state}}=2$, a depth-wise convolution window size $d_{\mathrm{conv}}=1$, and an MLP expansion factor of $\text{expand}=3$.

\subsection{Competing methods}\label{sec:baselines}
NeuroSSM was benchmarked against eight baselines that span traditional machine learning, classical DL, as well as recent transformer and SSM methods.

\textbf{SVM:}  
A linear $\ell_2$–regularized support–vector machine operating on static Pearson-correlation FC features \autocite{abraham2017deriving}. The cross-validated regularization weight was $C=1$.

\textbf{BrainNetCNN:}  
A convolutional model operating on FC features \autocite{kawahara2017brainnetcnn}. Cross-validated hyperparameters were $1\times10^{-4}$ learning rate.

\textbf{LSTM:}  
A recurrent model operating on raw BOLD signals, averaging hidden states across time for classification \autocite{dvornek2017identifying}. Cross-validated hyperparameters included 1 layer, 32 hidden units, 0.5 dropout, with 0.05 learning rate.

\textbf{STAGIN:}  
A graph-transformer model operating on sliding-window BOLD signals \autocite{kim2021learning}. Cross-validated hyperparameters were 4 model layers, 128 hidden units, a window size of 50, a stride of 3, and $1\times10^{-4}$ learning rate.

\textbf{BolT:}  
A transformer model employing fused local–window attention to process raw BOLD signals \autocite{bedel2023bolt}. The configuration included 4 transformer layers, 400 hidden units, a window size of 20, and a learning rate of $2\times10^{-4}$.

\textbf{FST-Mamba:}  
A hierarchical spatiotemporal state-space model operating on dynamic FC features \autocite{wei2025hierarchicalspatiotemporalstatespacemodeling}. The model used $d_{\mathrm{state}}=16$, $d_{\mathrm{conv}}=4$, $\mathrm{expand}=1$, with a learning rate of $1\times10^{-3}$.

\textbf{Mamba:}  
A state-space model built to operate on raw BOLD signals \autocite{mamba}. Cross-validated hyperparameters were $d_{\mathrm{state}}=2$, $d_{\mathrm{conv}}=2$, $\mathrm{expand}=3$, with a learning rate of $5\times10^{-4}$.

\section{Results}

\begin{table*}[t]
   \caption{Performance of competing methods on fMRI scans from HCP-Rest, HCP-Task, and PPMI. Accuracy (Acc.), F1 scores, and AUC metrics are reported as mean$\pm$std across all evaluated training-set fractions $S \in \{5,10,20,50,100\}\%$. Boldface denotes the best method and underlining denotes the second-best method for each metric in each dataset.}
\label{tab:comp_all_datasets}
  \centering
  \resizebox{\textwidth}{!}{%
  \begin{NiceTabular}{c|ccc|ccc|ccc}
  \toprule
   & \multicolumn{3}{c|}{\thead{HCP-Rest}} & \multicolumn{3}{c|}{\thead{HCP-Task}} & \multicolumn{3}{c}{\thead{PPMI}} \\
   \cmidrule(lr){2-4}\cmidrule(lr){5-7}\cmidrule(l){8-10}
   \thead{Method} & \thead{Acc.(\%)} & \thead{F1(\%)} & \thead{AUC(\%)} & \thead{Acc.(\%)} & \thead{F1(\%)} & \thead{AUC(\%)} & \thead{Acc.(\%)} & \thead{F1(\%)} & \thead{AUC(\%)} \\
  \midrule
  \thead{SVM} & \thead{75.20$\pm$4.78} & \thead{69.20$\pm$7.88} & \thead{83.23$\pm$5.97} & \thead{53.85$\pm$6.11} & \thead{52.50$\pm$5.57} & \thead{69.68$\pm$6.36} & \thead{83.22$\pm$4.68} & \thead{89.96$\pm$4.03} & \thead{60.28$\pm$18.36} \\
  \thead{BrainNetCNN} & \thead{72.05$\pm$3.20} & \thead{64.93$\pm$8.71} & \thead{77.13$\pm$3.49} & \thead{29.23$\pm$6.12} & \thead{24.45$\pm$6.62} & \thead{65.88$\pm$4.26} & \thead{82.26$\pm$5.59} & \thead{89.63$\pm$4.01} & \thead{52.67$\pm$18.23} \\
  \thead{LSTM} & \thead{73.35$\pm$2.97} & \thead{62.85$\pm$8.62} & \thead{83.03$\pm$2.86} & \thead{73.67$\pm$4.49} & \thead{72.43$\pm$5.15} & \thead{95.39$\pm$1.05} & \thead{82.78$\pm$6.66} & \thead{89.86$\pm$5.02} & \thead{61.70$\pm$17.59} \\
  \thead{STAGIN} & \thead{71.75$\pm$2.88} & \thead{65.10$\pm$5.40} & \thead{74.52$\pm$3.55} & \thead{40.53$\pm$2.93} & \thead{36.98$\pm$3.50} & \thead{74.12$\pm$2.06} & \thead{79.37$\pm$14.17} & \thead{85.41$\pm$14.28} & \thead{\underline{62.26$\pm$14.70}} \\
  \thead{BolT} & \thead{\underline{78.33$\pm$3.69}} & \thead{\underline{70.18$\pm$10.58}} & \thead{\underline{87.51$\pm$2.31}} & \thead{\underline{83.47$\pm$3.06}} & \thead{\underline{82.32$\pm$3.30}} & \thead{\underline{98.95$\pm$0.44}} & \thead{84.00$\pm$5.49} & \thead{90.43$\pm$4.94} & \thead{49.33$\pm$16.43} \\
  \thead{FST-Mamba} & \thead{67.89$\pm$3.09} & \thead{58.37$\pm$13.75} & \thead{76.03$\pm$3.41} & \thead{58.20$\pm$7.55} & \thead{55.87$\pm$8.63} & \thead{88.06$\pm$2.93} & \thead{77.22$\pm$11.88} & \thead{84.75$\pm$10.46} & \thead{58.87$\pm$13.12} \\
  \thead{Mamba} & \thead{68.87$\pm$5.41} & \thead{62.92$\pm$8.54} & \thead{75.34$\pm$7.53} & \thead{73.63$\pm$5.17} & \thead{72.52$\pm$5.46} & \thead{94.13$\pm$2.08} & \thead{\underline{84.87$\pm$4.47}} & \thead{\underline{91.50$\pm$2.95}} & \thead{59.87$\pm$9.03} \\
  \thead{NeuroSSM} & \thead{\textbf{81.76$\pm$4.02}} & \thead{\textbf{74.68$\pm$8.63}} & \thead{\textbf{89.81$\pm$2.02}} & \thead{\textbf{87.82$\pm$2.69}} & \thead{\textbf{87.24$\pm$2.48}} & \thead{\textbf{99.10$\pm$0.27}} & \thead{\textbf{86.43$\pm$3.56}} & \thead{\textbf{92.55$\pm$2.24}} & \thead{\textbf{67.47$\pm$17.75}} \\
  \bottomrule
  \end{NiceTabular}}
\end{table*}

\begin{figure*}[t]
  \centering
  \includegraphics[width=\linewidth]{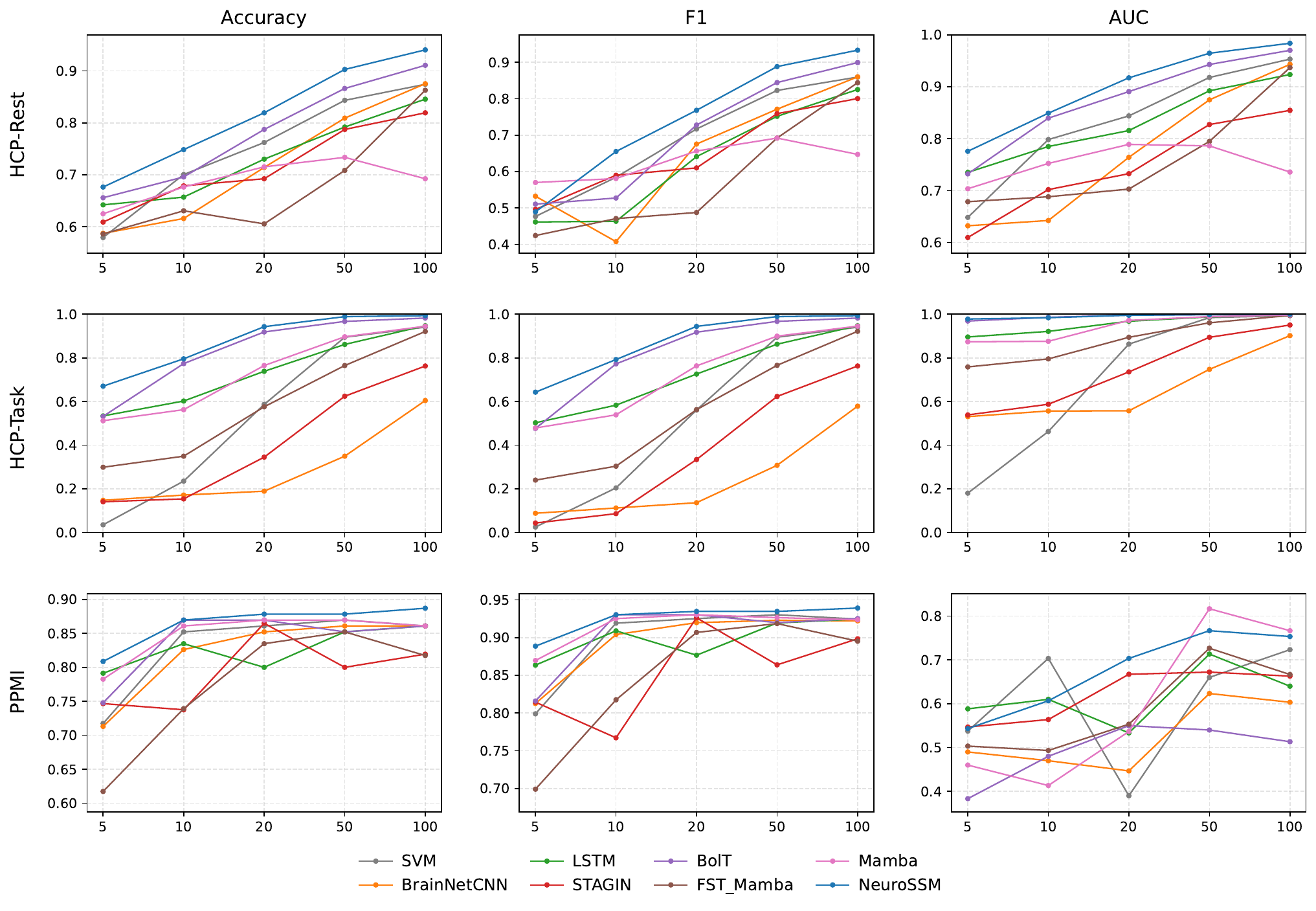}
  \caption{Learning efficiency curves on HCP-Rest, HCP-Task, and PPMI. Each panel displays accuracy, F1 score, or AUC measured on the held-out test set, for models trained on varying fractions $S \in \{5,10,20,50,100\}\%$ of subjects within the development set. The $x$-axis denotes $S$ (training-set fraction), and the $y$-axis shows the corresponding metric value.}
  \label{fig:learning_curves}
\end{figure*}

\subsection{Comparative demonstrations}
We evaluated NeuroSSM on three fMRI analysis tasks: gender detection on \textit{HCP-Rest}, cognitive task detection on \textit{HCP-Task}, and disease detection on \textit{PPMI}. Comparisons were performed against established baselines, including traditional (SVM), classical DL (BrainNetCNN, LSTM), transformer (STAGIN, BolT), and state-space (FST-Mamba, Mamba) models. To assess the learning efficacy of competing methods, we systematically varied the fraction of training subjects available on each dataset and examined model performance as a function of training set size $S \in \{5,10,20,50,100\}\%$. 

The learning curves for accuracy, F1, and AUC as a function of the training-set size are shown in Fig.~\ref{fig:learning_curves}. On HCP-Rest and HCP-Task, NeuroSSM consistently attains the strongest performance across $S$, with particularly notable advantages in the low-data regime (i.e., $S \leq 20\%$), indicating improved sample efficiency relative to transformer and SSM baselines. On PPMI, accuracy and F1 are near-saturated for most methods, whereas AUC exhibits substantially higher variability across $S$. NeuroSSM maintains competitive AUC overall and shows stronger ranking behavior at small-to-moderate training fractions, supporting its robustness in data-limited settings.

Aggregate performances across varying training-set fractions are summarized in Table~\ref{tab:comp_all_datasets}. On HCP-Rest, NeuroSSM achieves higher performance metrics in gender detection than all baselines ($p < 0.05$, signed-rank test). It improves over traditional baselines by $(+6.6\%,+5.5\%,+6.6\%)$, over classical DL by $(+9.1\%,+10.8\%,+9.7\%)$, over transformer baselines by $(+6.7\%,+7.0\%,+8.8\%)$, and over state-space baselines by $(+13.4\%,+14.0\%,+14.1\%)$ in (Acc, F1, AUC). On HCP-Task, NeuroSSM also yields the highest performance metrics, with benefits particularly evident for small to moderate training sets ($p<0.05$). NeuroSSM provides benefits of $(+34.0\%, +34.7\%, +29.4\%)$ over traditional baselines, $(+36.4\%, +38.8\%, +18.5\%)$ over classical DL, $(+25.8\%, +27.6\%, +12.6\%)$ over transformer, and $(+21.9\%, +23.0\%, +8.0\%)$ over state-space baselines. While all methods approach ceiling performance at large $S$, NeuroSSM maintains the highest AUC across $S$. Lastly, on PPMI, NeuroSSM yields the highest performance metrics in disease detection ($p<0.05$). It provides gains of $(+3.2\%,+2.6\%,+7.2\%)$ over traditional baselines, $(+3.9\%,+2.8\%,+10.3\%)$ over classical DL, $(+4.8\%,+4.6\%,+11.7\%)$ over transformer, and $(+5.4\%,+4.4\%,+8.1\%)$ over state-space baselines. While simpler linear baselines like SVM show competitive F1 scores, their weaker AUC reflects lower reliability, likely due to the higher data heterogeneity characteristic of PPMI. Taken together, these results suggest that NeuroSSM more effectively extracts subject-level latent structure from fMRI time series---capturing both intrinsic cognitive variables (resting-state gender and task-evoked cognition) and disease-related signatures---especially when supervision is limited.

\subsection{Ablation studies}\label{sec:ablation}
We conducted ablation studies to isolate the contributions of the multiscale processing and the differential input stream in NeuroSSM. We analyzed four variants: (i) a single-scale baseline without multiscale processing or differencing, (ii) a multi-scale-only variant, (iii) a differential-only variant, and (iv) the full model combining both components. Table~\ref{tab:ablation_NeuroSSM_streams_multi_dataset} lists results on the resting-state HCP-Rest and PPMI datasets, with results averaged across training-set fractions. Both multiscale processing and the differential stream improve performance over the single-scale baseline, while the full model achieves the strongest overall performance, indicating that the two components provide complementary benefits.

Next, we assessed the impact of the temporal resolution configuration $\boldsymbol{\tau}$ in the multiscale backbone (Table~\ref{tab:abl_tau_neurossm}). Overall, $\boldsymbol{\tau}=\{1,2,3\}$ and $\boldsymbol{\tau}=\{1,2,3,4\}$ yield closely matched performance across datasets. These findings corroborate that adding further scales ond the selected $\boldsymbol{\tau}=\{1,2,3\}$ configuration introduces additional model complexity without clear performance benefits.

\begin{table*}[t]
   \caption{Performance of NeuroSSM variants on resting-state fMRI scans from HCP-Rest, PPMI, and their average. Indicator columns specify whether the multi-scale stream, and the differential stream are included in the variant. Boldface denotes the best method and underlining denotes the second-best method for each metric in each dataset.}
\label{tab:ablation_NeuroSSM_streams_multi_dataset}
  \centering
  \resizebox{\textwidth}{!}{%
  \begin{NiceTabular}{cc|ccc|ccc|ccc}
  \toprule
   & & \multicolumn{3}{c|}{\thead{HCP-Rest}} & \multicolumn{3}{c|}{\thead{PPMI}} & \multicolumn{3}{c}{\thead{Average}} \\
   \cmidrule(lr){3-5}\cmidrule(lr){6-8}\cmidrule(l){9-11}
   \thead{Multi-scale} & \thead{Differential} & \thead{Acc.(\%)} & \thead{F1(\%)} & \thead{AUC(\%)} & \thead{Acc.(\%)} & \thead{F1(\%)} & \thead{AUC(\%)} & \thead{Acc.(\%)} & \thead{F1(\%)} & \thead{AUC(\%)} \\
  \midrule
  \thead{\xmark} & \thead{\xmark} & \thead{79.43$\pm$10.36} & \thead{71.36$\pm$16.93} & \thead{87.68$\pm$9.18} & \thead{86.09$\pm$1.74} & \thead{92.46$\pm$1.13} & \thead{49.87$\pm$7.87} & \thead{82.76$\pm$6.05} & \thead{81.91$\pm$9.03} & \thead{68.77$\pm$8.53} \\
  \thead{\cmark} & \thead{\xmark} & \thead{\underline{80.59$\pm$9.59}} & \thead{\underline{73.07$\pm$16.12}} & \thead{\underline{88.96$\pm$7.87}} & \thead{84.52$\pm$5.32} & \thead{91.14$\pm$3.99} & \thead{\underline{64.93$\pm$11.50}} & \thead{82.56$\pm$7.45} & \thead{82.11$\pm$10.06} & \thead{\underline{76.95$\pm$9.69}} \\
  \thead{\xmark} & \thead{\cmark} & \thead{80.05$\pm$10.61} & \thead{72.78$\pm$16.69} & \thead{88.11$\pm$9.32} & \thead{\underline{86.26$\pm$1.39}} & \thead{\textbf{92.58$\pm$0.88}} & \thead{52.53$\pm$7.67} & \thead{\underline{83.15$\pm$6.00}} & \thead{\underline{82.68$\pm$8.78}} & \thead{70.32$\pm$8.49} \\
  \thead{\cmark} & \thead{\cmark} & \thead{\textbf{81.76$\pm$9.70}} & \thead{\textbf{74.68$\pm$16.13}} & \thead{\textbf{89.81$\pm$7.68}} & \thead{\textbf{86.43$\pm$2.84}} & \thead{\underline{92.55$\pm$1.86}} & \thead{\textbf{67.47$\pm$8.64}} & \thead{\textbf{84.10$\pm$6.27}} & \thead{\textbf{83.61$\pm$9.00}} & \thead{\textbf{78.64$\pm$8.16}} \\
  \bottomrule
  \end{NiceTabular}}
\end{table*}

\begin{table*}[t]
   \caption{Performance of NeuroSSM variants on resting-state fMRI scans from HCP-Rest, PPMI, and their average. Variant models were built for varying resolution configurations $\boldsymbol{\tau}$. Boldface denotes the best method and underlining denotes the second-best method for each metric in each dataset.}
\label{tab:abl_tau_neurossm}
  \centering
  \resizebox{0.95\textwidth}{!}{%
  \begin{NiceTabular}{c|ccc|ccc|ccc}
  \toprule
   \thead{$\boldsymbol{\tau}$ set} & \multicolumn{3}{c|}{\thead{HCP-Rest}} & \multicolumn{3}{c|}{\thead{PPMI}} & \multicolumn{3}{c}{\thead{Average}} \\
   \cmidrule(lr){2-4}\cmidrule(lr){5-7}\cmidrule(l){8-10}
   \thead{} & \thead{Acc.(\%)} & \thead{F1(\%)} & \thead{AUC(\%)} & \thead{Acc.(\%)} & \thead{F1(\%)} & \thead{AUC(\%)} & \thead{Acc.(\%)} & \thead{F1(\%)} & \thead{AUC(\%)} \\
  \midrule
  \thead{$\{1\}$} & \thead{80.05$\pm$10.61} & \thead{72.78$\pm$16.69} & \thead{88.11$\pm$9.32} & \thead{\underline{86.26$\pm$1.39}} & \thead{\textbf{92.58$\pm$0.88}} & \thead{52.53$\pm$7.67} & \thead{83.15$\pm$6.00} & \thead{82.68$\pm$8.78} & \thead{70.32$\pm$8.49} \\
  \thead{$\{1,2\}$} & \thead{81.14$\pm$10.17} & \thead{74.48$\pm$15.77} & \thead{89.31$\pm$8.37} & \thead{86.26$\pm$2.36} & \thead{92.51$\pm$1.52} & \thead{64.73$\pm$10.37} & \thead{83.70$\pm$6.27} & \thead{83.49$\pm$8.64} & \thead{77.02$\pm$9.37} \\
  \thead{$\{1,2,3\}$} & \thead{\underline{81.76$\pm$9.70}} & \thead{74.68$\pm$16.13} & \thead{89.81$\pm$7.68} & \thead{\textbf{86.43$\pm$2.84}} & \thead{\underline{92.55$\pm$1.86}} & \thead{\textbf{67.47$\pm$8.64}} & \thead{\textbf{84.10$\pm$6.27}} & \thead{83.61$\pm$9.00} & \thead{\textbf{78.64$\pm$8.16}} \\
  \thead{$\{1,2,3,4\}$} & \thead{\textbf{81.90$\pm$9.11}} & \thead{\textbf{75.65$\pm$15.12}} & \thead{\textbf{89.98$\pm$7.36}} & \thead{85.91$\pm$3.45} & \thead{92.14$\pm$2.42} & \thead{66.33$\pm$10.28} & \thead{\underline{83.91$\pm$6.28}} & \thead{\textbf{83.90$\pm$8.77}} & \thead{78.15$\pm$8.82} \\
  \thead{$\{1,2,3,4,5\}$} & \thead{81.30$\pm$9.96} & \thead{\underline{74.86$\pm$16.40}} & \thead{\underline{89.81$\pm$7.42}} & \thead{86.26$\pm$2.29} & \thead{92.48$\pm$1.48} & \thead{\underline{67.47$\pm$8.92}} & \thead{83.78$\pm$6.13} & \thead{\underline{83.67$\pm$8.94}} & \thead{\underline{78.64$\pm$8.17}} \\
  \bottomrule
  \end{NiceTabular}}
\end{table*}

\section{Discussion} 
We introduced NeuroSSM, a novel selective state-space architecture designed to capture hierarchical and context-aware representations in fMRI time series for downstream analysis. By processing BOLD signals at various temporal scales while simultaneously injecting first-order derivative information, NeuroSSM integrates long-range context with a computational cost that remains linear in sequence length (i.e., scan duration), avoiding the quadratic complexity of attention mechanisms in transformers. Classification is then performed on a pooled representation of these rich spatiotemporal features. Demonstrations on resting-state and task-based fMRI datasets show that NeuroSSM achieves superior performance compared to state-of-the-art baselines, including convolutional, recurrent, graph-based, transformer, as well as recent SSM models.

Our primary focus in this study was on classification models with categorical outputs for gender, cognitive task, and disease detection. The global pooling over the final hidden features provides a condensed, high-level representation of the entire fMRI scan. However, brain activity also encodes information about continuous variables \autocite{ccukur2013attention}. NeuroSSM is readily suitable for analyzing cortical representations of such features. The final pooled representation vector can be coupled with a linear output layer or a multi-layer perceptron with a linear activation function, enabling NeuroSSM to decode continuous variables from fMRI scans.

A common standard in computational neuroimaging is to build decoding models that predict external stimuli or task conditions from measured brain activity \autocite{laconte2011decoding, andersson2011real}, a framework we followed here. An alternative approach involves building encoding models that predict brain activations from stimulus or task features \autocite{nishimoto2021modeling, celik2021cortical, shahdloo2022task, anderson2016representational, ngo2022transformer}. For cognitive neuroimaging studies, the experimental time course of a task could serve as input to a NeuroSSM-based architecture, which could then be trained to predict the measured BOLD responses on a region-by-region basis. Assessing the effectiveness of NeuroSSM in this neural encoding framework, particularly through the use of multi-scale linking where finer-scale features are conditioned on coarser representations to ensure consistent context propagation \autocite{kabas2025physicsdrivenautoregressivestatespace}, remains an important direction for future work.

Our analysis of Parkinson's disease on the PPMI dataset relied solely on information from resting-state fMRI scans. While our results are promising, studies suggest that disease detection can be enhanced by incorporating auxiliary information, such as patient demographics or clinical assessments \autocite{dvornek2018combining}. Furthermore, many neurological disorders, including Parkinson's, present with distinct signatures in other imaging modalities like structural or diffusion-weighted MRI \autocite{pyatigorskaya2014review}. NeuroSSM's performance could be further improved by integrating such multi-modal data. Auxiliary information could be fused with the latent representation just before the final classification layer, while additional imaging modalities could be processed by parallel encoder streams whose outputs are concatenated with the fMRI features.

As is common practice in neuroimaging, we parcellated each subject's brain using a standard atlas in a template space and fed the average BOLD responses from these ROIs into our model. While this ensures consistent ROI definitions across subjects, the spatial registration to a common template can be a lossy transformation. Such losses might be mitigated by defining ROIs in each subject's native brain space. This could be achieved by back-projecting the template ROIs using the inverse of the registration transform. Alternatively, a 3D CNN could be integrated as a spatial front end to NeuroSSM, learning to extract features from the volumetric fMRI data before the state-space backbone models their temporal evolution \autocite{nguyen2020attend, malkiel2021pre, choi2025neuromamba}. In this context, utilizing specialized SSM blocks that independently capture spatial and channel-wise contextual dependencies may further improve representational power when processing high-dimensional volumetric data \autocite{ozturk2025denomamba}.

We trained all competing models from scratch on datasets containing several hundred subjects. Despite training all competing models from scratch, our results on the HCP-Task learning curves suggest that NeuroSSM performs robustly even when trained with a limited number of subjects. For applications where only compact datasets are available, its performance could be further bolstered via transfer learning \autocite{devlin2018bert, dalmaz2021resvit}. For instance, NeuroSSM could be pre-trained on a large-scale public dataset like HCP and then fine-tuned on a smaller, specific cohort. Reliable augmentation via image synthesis based on advanced procedures can also help alleviate data scarcity \autocite{dar2022adaptive, ozbey2022unsupervised, chen2025coupling, atli2025i2imambamultimodalmedicalimage}. Other strategies like federated learning could facilitate training on diverse, multi-institutional datasets without compromising patient privacy, further enhancing the generalization capabilities of the model \autocite{elmas2022federated, dalmaz2022one}. A systematic exploration of the data efficiency of NeuroSSM remains an important topic for future research.

\section{Conclusion}
In this study, we introduced NeuroSSM, a novel selective state-space architecture for fMRI classification that operates efficiently on raw BOLD sequences. It uses a multiscale backbone to capture varied temporal dynamics and a parallel differential stream to enhance sensitivity to transient activity, all with linear computational cost. Validated on large-scale HCP and PPMI datasets, NeuroSSM achieved state-of-the-art accuracy in gender, disease (Parkinson's), and cognitive task detection, surpassing existing baselines. NeuroSSM provides an efficient and sensitive framework for fMRI analysis, which is well-suited for future clinical and cognitive extensions, including the decoding of continuous variables and the integration of multimodal data.

\section*{Acknowledgements}
This study was supported in part by a TUBITAK BIDEB scholarship awarded to F. Gen\c{c}, and by a TUBA GEBIP 2015 fellowship, a BAGEP 2017 fellowship, and a TUBITAK 121N029 grant awarded to T. Çukur. \revhl{Funds have been provided by the European Joint Programme Neurodegenerative Disease Research (JPND) 2020 call “Novel imaging and brain stimulation methods and technologies related to Neurodegenerative Diseases” for the Neuripides project ‘Neurofeedback for self-stimulation of the brain as therapy for Parkinson Disease’. The Neuripides project has received funding from the following funding organizations under the aegis of JPND: The Netherlands, The Netherlands Organization for Health Research and Development (ZonMw); Germany, Federal Ministry of Education and Research (BMBF); Czech Republic, Ministry of Education, Youth and Sports (MEYS); France, French National Research Agency (ANR); Canada, Canadian Institutes of Health Research (CIHR); Turkey, Scientific and Technological Research Council of Turkey (TUBITAK).}

\footnotesize
\printbibliography{}

\end{document}